\documentclass{ws-ijmpe}
\usepackage[super,compress]{cite}
\usepackage{graphics}
\usepackage{graphicx,colordvi}

\expandafter\ifx\csname package@font\endcsname\relax\else
\expandafter\expandafter
\expandafter\usepackage
\expandafter{\csname package@font\endcsname}
\fi
\DeclareRobustCommand\substyle{\name@idx{document substyle}}
\DeclareRobustCommand\classoption{\name@idx{document class option}}
\DeclareRobustCommand\classname{\name@idx{document class}}
\def\name@idx#1#2{{\ttfamily#2}
\index{#2\space#1=\string\ttt{#2}\space#1}\index{#1>#2=\string\ttt{#2}}}

\newcommand{\beq}[0]{\begin{equation}}
\newcommand{\eeq}[0]{\end{equation}}
\newcommand{\bea}[0]{\begin{eqnarray}}
\newcommand{\eea}[0]{\end{eqnarray}}
\def\r{{\bf r}}
\def\p{{\bf p}}

\def\d{\hbox{d}}
\def\be{\begin{equation}}
\def\ee{\end{equation}}
\def\bea{\begin{eqnarray}}
\def\eea{\end{eqnarray}}
\def\l{\label}

\def\om{\omega}

\def\hahat{\hat{H}}

\def\hahat0{\hat{H}_0}

\def\cos{\hbox{cos}}

\def\exp{\hbox{exp}}

\def\Im{{\mbox {\rm Im}}}
\def\Re{{\mbox {\rm Re}}}

\def\vareps{\varepsilon}

\def\siml{\hbox{\kern.1em \lower.6ex \hbox{$\sim$} \kern-1.12em
          \raise.6ex \hbox{$<$} \kern.1em }}
\def\simg{\hbox{\kern.1em \lower.6ex \hbox{$\sim$} \kern-1.12em
          \raise.6ex \hbox{$>$} \kern.1em }}

\def\siml{\hbox{\kern.1em \lower.6ex \hbox{$\sim$} \kern-1.12em
 \raise.6ex \hbox{$<$} \kern.1em}}
\def\simg{\hbox{\kern.1em \lower.6ex \hbox{$\sim$} \kern-1.12em
 \raise.6ex \hbox{$>$} \kern.1em}}

\newcommand{\beqar}{\begin{eqnarray}}
\newcommand{\eeqar}[1]{\label{#1} \end{eqnarray}}

\begin{document}

\markboth{D.V.~Gorpinchenko, A.G.~Magner, J.~Bartel}{Semiclassical and quantum shell-structure
  calculations
 of the moment of inertia}

\catchline{}{}{}{}{}

\title{Semiclassical and quantum shell-structure calculations
 of the moment of inertia}

\author{D.V.~Gorpinchenko}

\address{{\it Institute for Nuclear Research, Prospekt Nauki 47,
    03028 Kyiv, Ukraine}\\
first\_dimag1991@gmail.com}

\author{A.G.~Magner}

\address{{\it Institute for Nuclear Research, Prospekt Nauki 47,
    03028 Kyiv, Ukraine}\\
  second\_magner@kinr.kiev.ua}

\author{J.~Bartel}

\address{{\it Institut Pluridisciplinaire Hubert Curien, CNRS/IN2P3,
 Université de Strasbourg, 23 rue du Loess, 67000 Strasbourg, France}\\
third\_Johann.Bartel@iphc.cnrs.fr}

\maketitle

\begin{history}
\received{Day Month Year}
\revised{Day Month Year}
\end{history}

\begin{abstract}
Shell corrections to the moment of inertia (MI) are calculated for a
Woods-Saxon potential of spheroidal shape and at different deformations. 
This model 
potential is chosen to have a 
large depth and a small surface diffuseness 
which makes it resemble the analytically solved spheroidal cavity in the 
semiclassical approximation.
For the consistent statistical-equilibrium collective rotations under 
consideration here, the MI is obtained within the cranking model in an 
approach which goes beyond the quantum perturbation approximation based 
on the non perturbative energy spectrum, and is therefore applicable to 
much higher angular momenta.
For the calculation of the MI shell corrections $\delta \Theta$, the 
Strutinsky smoothing procedure is used to obtain the average occupation 
numbers of the particle density generated by the resolution of the 
Woods-Saxon eigenvalue problem. 
One finds that the major-shell structure of $\delta \Theta\,$, as 
determined in the adiabatic approximation, is rooted, for large as well 
as for small surface deformations, in the same inhomogenuity of the 
distribution of single-particle states near the Fermi surface as the 
energy shell corrections $\delta E$.
This fundamental property is in agreement with the semiclassical results 
$\delta \Theta \propto \delta E$ obtained analytically within the non 
perturbative periodic orbit theory for any potential well,
in particular
for the spheroidal cavity, and for any deformation, even 
for large deformations where bifurcations of the equatorial orbits play a 
substantial role. Since the adiabatic approximation, $\omega \ll \Omega$, 
with $\hbar \Omega $ the distance between major nuclear shells, is 
easily obeyed even for large angular momenta typical for  high-spin 
physics at large particle numbers, our model approach seems to represent 
a tool that could, indeed, be very useful for the description
of such nuclear systems.
\end{abstract}


\ccode{PACS numbers: 21.10. Ev, 21.60. Cs, 24.10 Pa}


\section{Introduction}
\label{sec-introd}

Many significant phenomena in nuclear rotations can be explained within 
a theoretical approach based on the cranking model
\cite{In54,BM75,RS80},
and on the Strutinsky shell-correction 
method (SCM) \cite{St67,BD72}.
This approach was extended by 
Pashkevich and Frauendorf \cite{PF75} 
to the description of collective rotational bands. 
For a deeper understanding of the correspondence between the classical 
and the quantum approach and their application to high-spin physics, 
it is worthwhile to analyze the shell components of the MI within the 
semiclassical periodic-orbit theory (POT) 
\cite{Gu71,Gu90,SM76,SM77,BB03,MY11}.
The classical perturbation expansion \cite{Cr96} of 
the action integral as function of the rotation frequency has been used
in Ref.\ \refcite{DF04} in the POT calculation of the MI shell 
corrections for a spheroidal cavity mean field.
For the non perturbative semiclassical POT description 
 of adiabatic rotations
 in the improved stationary phase method\cite{MA02,MY11} (ISPM) 
   extended to the rotational
   phenomena (ISPMR) \cite{MS10,MG14,GM15,MG17},
the rotational angular frequency $\hbar \omega$
has to be smaller by about an order of the magnitude than
the distance $\hbar \Omega$ between major shells near the Fermi surface.
The condition $\omega \ll \Omega$ therefore allows for much larger
angular momenta than those required in the quantum perturbation approach
of the MI Inglis cranking formula
  where $\hbar \omega$ is required to be smaller than the 
  distance between neighboring single-particle (s.p.) states,
  as explained below.
Our ISPM approach 
has been extended,
to the bifurcation phenomena at large deformations 
\cite{MA02,MY11}, and was applied within the
cranking model to the description of 
collective rotations (around an axis perpendicular to the symmetry axis)
in a harmonic-oscillator \cite{MS10} and a spheroidal cavity \cite{GM15,MG17}.
The fact that the ISPM is 
  also
working well 
  even
for partially non-integrable 
systems such as the Woods-Saxon potential, as will be shown below, causes 
the ISPMR to be well adapted to describe nuclear systems at large 
deformations and at high angular frequencies (high-spin physics).
                                                                     \\[ -2.0ex]

It appears only natural to use the semiclassical POT in connection
with the cranking model which is, strictly speaking 
of semiclassical nature because the 
collective rotation of the nuclear many-body systems is described there
as a classical transformation from the laboratory to the body-fixed coordinate 
system, rotating around the former with fixed angular velocity\cite{BM75,RS80}, 
thus defining angular momentum and rotation angles simultaneously.
Using this semiclassical picture, one can reduce the complex problem of 
the rotation of a many-body system to a much simpler diagonalization of 
an effective mean-field Hamiltonian in the rotating frame.
In the semiclassical limit of large particle numbers $N$,
one can neglect the energy 
distances between nearest-neighbor quantum states. 
In the semiclassical approximation 
the energy spectrum of the unperturbed Hamiltonian 
(in the absence of rotations) can be considered 
as quasi-continuous, and, therefore, 
any finite rotational excitation could never be considered to be 
small as compared to the distance $D$ between non perturbative neighboring 
levels, thus violating the condition of applicability of the
quantum perturbation 
expansion (QPE). 
It should be noted, nonetheless, that the quantum pairing theory has been successful 
   in describing rotating nuclear systems through the Inglis-Valatin cranking formula \cite{NP61,SG89,TH95}, 
   where pairing correlations produce in the spectrum a considerable gap, which often is much larger than $D$. 
Since nuclei are finite Fermi systems, 
the correlations beyond the 
mean field smear out the sharp 
Mottelson-Valatin phase transitions \cite{MV60} between 
rotational states obtained with and without accounting for the pairing correlations,
an effect which is of great importance,
especially for super-deformed nuclei (see, e.g., Refs. \refcite{SG89,TH95}).
Pairing correlations are certainly decreasing
for the case of excited (finite temperature) nuclei, but 
correlations beyond the mean field and thermal excitations again
smear out such a transition \cite{EE93}. 
The description of excited nuclei within our, at the present moment, somewhat academic model, 
taking, in addition, pairing correlations into account, goes far beyond the scope of the present work which is rather
to be understood as a preliminary step to a full description of such nuclear systems.
The simple Inglis cranking formula \cite{In54}, without taking into 
account correlations beyond the mean field, 
being founded on stationary perturbation theory, clearly has its limitations
which causes this approximation to simply fail whenever the conditions of
its application are not fulfilled, as this is the case in the semiclassical approach.
                                                                    \\[ -2.0ex]

The ISPM is found to be the faster converging, the larger the particle numbers
$N$ (Refs.~\citen{MA02,MY11}).
Within the semiclassical POT, the coordinate representation of the MI
through the Green's function $G$ (Refs.~\citen{MG14,GM15,MG17}) 
is extremely useful, since it allows to weaken the 
applicability criterion of the QPE approximation,
because the leading 
order 
terms of the POT in 
the expansion over the small parameter
$\hbar/S \sim N^{-1/3}$ (where $S$ is the
action integral for the dominating POs) 
is related to a statistical averaging
over many quantum states. Therefore, the
maximal rotational excitation energy $\hbar \omega$ for which 
  this 
approximation is valid becomes now significantly larger than the 
nearest-neighbor s.p.\ level spacing $D$ around the Fermi surface 
$\vareps^{}_F$. 
At the same time, $\hbar \omega$
still remains somewhat smaller than the 
energy distance $\hbar \Omega$ between major shells. 
These two conditions are in contrast to the energy spectrum
representation of the MI QPE approach where,
in the derivation of the 
standard Inglis cranking formula the excitation energies are required 
to be small with respect to the s.p.\ level spacing $D$.
This more severe restriction is obviously avoided by using the
coordinate-space representation of the Green's 
function $G$ in the determination of the MI \cite{GM15}.
                                                                    \\[ -2.0ex]

In contrast to the QPE approach of Ref.\ \refcite{PF75}, we will therefore avoid to use the 
well-known Inglis cranking formula \cite{In54} based on the QPE, but shall use instead
another approach based on the concept of a statistical 
equilibrium rotation \cite{LLv5} with a
    {\it generalized rigid-body} (GRB) moment of 
inertia $\Theta_{\rm GRB}$ 
(see also Refs.\ \citen{MG17,GM15,MS10})
\be\l{MIGRB}
  \Theta \approx \Theta_{\rm GRB}=m\int \d\r~ r_{\perp}^2 \; \rho(\r)~,
\ee
which is found to constitute the main  contribution to the moment of 
inertia at  
nuclear equilibrium deformations.
Here $m$ is the nucleon mass and 
$r_{\perp}$ the distance between a given point 
$\r$ of the nucleus and the rotation axis. 
Since $\rho(\r)$ is here the full one-body quantum particle-number 
density, this MI expression goes beyond the classical (rigid body) MI.
According to the SCM \cite{St67,BD72,PF75}, one has 
$\rho(\r) = \tilde{\rho}(\r) + \delta \rho(\r)$
in Eq.\ (\ref{MIGRB}), where $\tilde{\rho}$
is a smooth density and $\delta \rho(\r)$ its shell correction. 
It is obviously this shell component $\delta \rho(\r)$ which determines the 
MI shell correction $\delta\Theta \approx \delta \Theta_{\rm GRB}$ at the equilibrium deformation. 
To avoid any possible misunderstanding that could arise by confusing 
the quantum MI $\Theta_{\rm GRB}$ with its classical (rigid body) counterpart 
that would be defined through the classical density of the Thomas-Fermi model, 
we introduced the GRB moment of inertia $\Theta_{\rm GRB}$
of Eq.\ (\ref{MIGRB}) through the quantum particle density $\rho(\r)$ 
(see Refs.\ \citen{MG17,GM15,MS10}).
The concept of a statistical equilibrium rotation
implies in fact a consistency condition in the sense that the 
average angular momentum is just the one of the collective rotation 
of the nuclear body in the absence of any
inconsistent contribution, coming e.g.\ from s.p.\
motion, as this has been explained in detail 
for the case of a harmonic oscillator mean field in Refs.\ \citen{BM75}
and \citen{MS10}.
Eq.\ (\ref{MIGRB}) that is statistically accounting for the 
consistent collective rotational motion, even in cases where the QPE is
no longer applicable, is in fact also 
perfectly suited to be used in the framework of the ISPMR with a semiclassical particle density $\rho$, which is the 
ISPM sum of a smooth part $\tilde{\rho}$ and a shell component $\delta \rho$.
                                                                    \\[ -2.0ex]

For adiabatic collective rotations (rotations at statistical equilibrium) 
the MI is 
 thus
described as the ISPMR sum of a smooth Extended
Thomas-Fermi (ETF) MI $\Theta^{}_{\rm ETF}$ (Ref.~\refcite{BQ94})
and shell corrections $\delta \Theta$ (Refs.~\citen{MS10,GM15,MG17}).
In a more realistic description of 
the MI for collective rotations, the ETF approach has already been successful, 
as in the case of the nuclear energy 
by including 
self-consistency and spin effects into the calculations, 
which would be especially important 
when used in connection with nuclear systems of large deformation 
and/or large rotation frequencies \cite{BQ94,MG17}.
                                                                    \\[ -2.0ex]

It has been shown within the ISPMR \cite{MS10,GM15,MG17},
that one can obtain, through a semiclassical 
phase-space trace formula, a quite accurate approximate analytical 
expression for the MI shell components $\delta \Theta$ in terms of the 
energy shell corrections $\delta E$, 
  for an arbitrary potential well,
a relation which has been worked out for integrable Hamiltonians, such as 
for a harmonic oscillator\cite{MS10} or a spheroidal-cavity\cite{GM15,MG17}
mean field.   
As demonstrated in Refs.\  \refcite{GM15} and 
\refcite{MG17}, this relation has been
derived \cite{GM15} within the ISPMR 
including equatorial orbits, which are important, 
especially for large deformations 
because of bifurcations which then become crucial \cite{MA02,MS10,MY11,GM15,MG17}.
In the following we are going to compare the quantum MI shell 
corrections $\delta \Theta$ with the energy shell corrections $\delta E$, 
calculated both by the SCM, 
for a deformed Woods-Saxon (WS) potential with large depth and small surface 
diffuseness (a system already studied 
in Ref.\ \refcite{SM77}) at different spheroidal deformations. 
It should, however, be made clear from the very beginning 
that one is not 
going to obtain for the here considered WS mean-field potential the same 
quality of correspondence between energy and MI shell components as this has 
been the case for the harmonic-oscillator potential well \cite{MS10}.
This is mainly due to the fact that the harmonic oscillator Hamiltonian is 
one of the very few cases (like the free particle) where
the stationary-phase method yields an {\it exact} result \cite{BB03}.
This is certainly not the case for the partially non-integrable WS potential
and one is not going to expect a strict proportionality between energy and 
MI shell corrections as predicted on the semiclassical level
by Ref.~\refcite{GM15}.  
The best one can hope for is that such a correspondence can be established 
on the level of major shell or major sub-shell structures.
                                                                    \\[ -2.0ex]

In Sect.\ \ref{sec-mishellcor} the essential elements of our 
approach will be presented, including the standard 
cranking model formulation (Subsec.\ \ref{subsec-cranmod}),
and the basic concept of a consistent 
statistically equilibrium rotation (Subsec.\ \ref{subsec-stateqrot}).
Sections \ref{sec-qm} and \ref{sec-scl} are devoted respectively to the 
quantum and the   semiclassical ISPMR resolution  
of the MI calculation. 
Our results are discussed in Sec.\ \ref{sec-comp} and a short summary and 
outlook on possible extensions of our approach are given in Sec.\ 
\ref{sec-concl}. Some details of the quantum calculation are presented 
in the Appendix.
\section{MI shell corrections}
\l{sec-mishellcor}

\subsection{The cranking model}
\l{subsec-cranmod}

Before explaining in some detail the cranking model which we use in our approach, 
   let us mention that, even though it is of semiclassical nature, as explained in the introduction, 
   it is not just a classical approximation, but can, in fact, be derived in a full quantum-mechanical
   approach \cite{BM70} through a variational principle from a variation after projection onto a 
   good angular momentum, under the conditions of large system deformations and large particle numbers.
Within the standard cranking model \cite{In54},
collective rotations of a Fermi system associated with a many-body 
Hamiltonian can be described in the
mean-field
approximation. 
The complex problem of a rotating many-body Fermi 
system can then be reduced, in the restricted subspace of Slater 
determinants, to a much simpler eigenvalue problem of a s.p.\ Hamiltonian
\be\l{Routhian}
  \hat{\cal H}^{\om} = \hat{\cal H}
                         - \boldsymbol{\om} \cdot \hat{\boldsymbol{\ell}} 
                     = \hat{\cal H} - \om \cdot \hat{\ell}_x \;, 
\ee
where $\hat{\boldsymbol \ell}$ is the s.p.\ angular-momentum operator with component 
$\hat{\ell}_x$, having defined 0$x$ as the 
rotation axis perpendicular to the
symmetry 0$z$ axis.
The Hamiltonian (\ref{Routhian}) is
usually referred to as the {\it Routhian}.
For simplicity,
we shall discard the spin and isospin degrees of freedom, 
in particular
the spin-orbit and asymmetry interactions.
The rotation frequency $\omega$ of the body-fixed coordinate system with 
respect to the laboratory frame is the Lagrange multiplier of our problem,
associated with the constraint on the nuclear angular momentum $I_x$.
The angular velocity $\omega$  needs to be adjusted in 
such a way that the 
quantum average $\langle \hat{\ell}_x \rangle^{\om}$ of the s.p.\ orbital 
angular-momentum operator $\hat{\ell}_x$ yields the required angular momentum 
$I_x$. 
This quantum average is obtained in a similar way as
the expectation value  
of the many-body Routhian
(\ref{Routhian})
in the subspace of Slater determinants,
\be\l{constraint0}
  \langle \hat{\ell}_x \rangle^{\om} \equiv d_s \sum_i n_i^{\om} 
  \int d \r \;\psi_i^{\om}\!\left(\r\right) \; \hat{\ell}_x\;
  \overline{\psi}_i^{\om}\!\!\left(\r\right) = I_x\;,
\ee
where $d_s$ is the spin (spin-isospin) degeneracy of the s.p.\ states
$| \psi_i^{\om} \rangle$,
$n_i^{\om}$ their occupation numbers, with corresponding eigenvalues 
$\vareps_i^{\om}$ and eigenfunctions $\psi_i^{\om}(\r)$ of
$\hat{H}^{\om}$, 
Eq.\ (\ref{Routhian}), and $\overline{\psi}_i^{\om}(\r)$ their complex 
conjugate.   
For relatively small angular velocities $\om$ and at zero nuclear  
temperature, the chemical potential $\lambda^{\om}$ is, to a good 
approximation, equal to the Fermi energy:
$\lambda^{\om} \approx \vareps_{{\rm F}} = \hbar^2 k_{\rm F}^2/2 m$,
where $\hbar k_{{\rm F}}$ is the Fermi momentum. Within
the same approach,
one approximately has for the particle number
\bea\l{partconspert}
  \hspace{-0.3cm}
  N = d_s\sum_i n_i^{\om}
  \int \d \r\; \psi_i^{\om}(\r)\;\overline{\psi}_i^{\om}(\r)
 = d_s \int_{0}^{\infty} \d \vareps \; n_{}^{\om}(\vareps) \,,
\eea
where the occupation numbers $n^{\omega}(\varepsilon)$ depend on the
chemical potential $\lambda^{\om}$ which has to be adjusted to obtain the 
desired particle number $N$.

The collective MI $\Theta_x$ for a rotation around the $x$ axis 
can be considered as the 
response of the quantum average
$\delta \langle \hat{\ell}_x \rangle^{\om}$ to the
external cranking field $-\omega \hat{\ell}_x$
in Eq.\ (\ref{Routhian}) \cite{MS10,MG14,GM15,MG17,MV91},
\be\l{response}
\delta \langle \hat{\ell}_x \rangle^{\om}=
\Theta_x \delta\om\;,
\ee
where
\be\l{thetaxdef}
  \Theta_{x} = \frac{\partial \langle \hat{\ell}_x\rangle^\om}{\partial \om}
             =\frac{\partial^2 E(\om)}{\partial \om^2}\;,
\ee
with
\be\l{Eomeg}
E(\om) = \langle \hat{\cal H}^\om \rangle
\approx E(0) + \frac{1}{2} \Theta_x \omega^2
\approx E(0) + \frac{I^{2}_x}{2 \Theta_x} \; .
\ee
For a nuclear rotation around the $x$ axis, one can treat, as shown in
Refs.\ \citen{In54,BM75,PF75}, the term $-\om \, \hat{\ell}^{}_x$
as a small perturbation.
With the constraint (\ref{constraint0})
and the MI, Eq.\ (\ref{thetaxdef}), if 
treated in 
second-order quantum perturbation theory,
one obtains the well-known Inglis cranking formula \cite{In54,BM75,RS80}. 
                                                                    \\[ -2.0ex]

Thus the collective rotational moment of inertia $\Theta_x$, that we are talking about, can be understood 
   within the cranking model as a response functions, describing the response of the nuclear system under the 
   influence of the external rotational velocity field $-\omega \, \hat{\ell}^{}_x$. 
   It is, indeed, well known that nuclei are Fermi 
   liquids and that such a response function depends in a sensitive way on the properties of the system around 
   the Fermi surface. Pairing correlations play an important role in this context. Cranking moments of inertia 
   calculated without taking pairing correlations into account are, however, well known to overestimate the 
   experimentally observed MI values substantially, in particular in well deformed nuclei. One could argue that
   the pairing correlations are weakened by the rotational motion \cite{MV60}, what is known as the Coriolis 
   anti pairing (CAP) effect (see also Ref.\ \citen{QB20}). 
   It remains, however, that neglecting pairing correlations completely, as we are
   doing in our approach, makes that it represents, in a way, simply an academic problem, which is, as already 
   pointed out in the introduction, to 
   be understood as a first step to pave the way for a more complete theory.
                                                                    \\[ -2.0ex]
   
For the derivation of the MI shell corrections 
within the SCM \cite{St67,PF75}, beyond the QPE approach,
it turns out to be helpful to use the
coordinate-space
representation of the MI
through
the one of
    the s.p.\ Green's functions $G\left(\r_1,\r_2;\vareps \right)$
as this was done for the other transport coefficients 
      in Refs.\ \citen{MV91,MG17}.
Taking advantage of the analogy of our problem of a rotating many-body system 
with magnetism, where the magnetization ${\boldsymbol M}$ is proportional to 
the field strength ${\boldsymbol B}$ with the magnetic susceptibi\-lity $\chi$
as the proportionality constant \cite{RU96,FK98},
the MI $\Theta_x$,
Eq.\ (\ref{thetaxdef}),
can be expressed in a coordinate representation, as a 
      kind of
susceptibility, or as 
the response function for collective vibrations \cite{MG17}, 
in terms of the  
Green's function $G$ (see also Refs.\ \citen{MV91,GM15}).
For adiabatic rotations, one then has
\bea\label{micoorrep}
\Theta_{x}&=&\frac{2 d_s}{\pi}\int^{\infty}_0 {\rm d}\vareps\;
n({\vareps}) 
\int {\rm d} \r_1 \int {\rm d} \r_2 \; 
\ell_{x}(\r_1)\; \ell_{x}(\r_2)  \nonumber\\
&\times&  \Re \left[G\left(\r_1,\r_2;\vareps \right) \right]\;
\Im \left[G\left(\r_1,\r_2;\vareps\right) \right]\; ,  
\eea
where 
$\boldsymbol \ell =[\r \times \p]$ is the particle angular momentum.
Formally, with the help of the spectral
representation of the Green's function $G\left(\r_1,\r_2;\vareps \right)$, 
one can also obtain from Eq.~(\ref{micoorrep}) the famous
Inglis formula for the MI \cite{BM75,RS80}.
                                                                    \\[ -2.0ex]

 \subsection{Statistically equilibrium rotation}
\l{subsec-stateqrot}

For a semiclassical statistical-equilibrium rotation with  
constant frequency $\om$, 
one approximately obtains\cite{MS10,MG17}
Eq.~(\ref{MIGRB}) 
for the MI $\Theta_x$ in terms of the GRB MI
(to simplify the notation, the sub-script $x$ will be omitted in what follows),
\be\l{misplit}
  \Theta \approx m \int \d \r \; r_{\perp}^2 \, \rho(\r) 
         = \widetilde{\Theta} + \delta \Theta\;, 
\ee
with $r_{\perp}^2 \!=\! y^2 \!+\! z^2$
and the smooth part 
$\widetilde{\Theta}=m \int \d \r \; r_{\perp}^2 \,\tilde{\rho}(\r) $ 
of the MI\cite{BQ94}, while the shell correction
is given by\cite{MS10,GM15,MG17}
\be\l{dtsclsplit}
 \delta \Theta = 
 m \int {\rm d}\r \; r_{\perp}^2 \, 
 \delta \rho(\r)\;,
\ee
where Eq.~(\ref{misplit}) is a local approximation (valid for 
statistically averaged rotations \cite{MS10,MG14,GM15}) to the 
more general equation (\ref{micoorrep}).
                                                                    \\[ -2.0ex]

The separation in Eq.~(\ref{misplit}) of the MI into a smooth average part and a 
shell correction has, of course, its origin in the corresponding  
      subdivision
of the spatial particle density 
\be\l{partdensplit} 
 \rho(\r)\! = 
 \! - \frac{1}{\pi}\Im \!\int\! {\rm d} \vareps\; n\left(\vareps\right)
 \left[G\left(\r_1,\!\r_2;\vareps\right)\right]_{\r_1=\r_2=\r}\!
 \!=\! \widetilde{\rho} \!+\! \delta\! \rho
\ee
into a smooth part $\widetilde{\rho}(\r)$ and a shell correction 
\be\l{drhoscl}
 \delta \rho\left(\r\right) \!=\!
 -\frac{1}{\pi} \Im  \int \d\vareps\; \delta n\left(\vareps\right) \nonumber\\
 \!\left[G\left(\r_1,\!\r_2;\vareps\right)\right]^{}_{\r_1=\r_2=\r}\; ,
\ee
where $G\left(\r_1,\!\r_2;\vareps\right)$ is the one-body Green's function.
Eq.~(\ref{partdensplit}) is stemming originally from the 
standard decomposition of the occupation numbers 
into smooth and fluctuating (shell) parts as usual in the SCM \cite{BD72}
\begin{equation}\label{ndecomp}
 n = \widetilde{n} + \delta n \; .
\end{equation}
Notice that, for the example of a harmonic oscillator potential, the GRB MI, 
Eqs.~(\ref{MIGRB}) and (\ref{misplit}), are identically obtained under the condition 
\begin{equation}\label{SelfCC}
  \mbox{N}_x \, \omega_x = \mbox{N}_y \, \omega_y = \mbox{N}_z \, \omega_z,\qquad
  \mbox{N}_i = \sum_\kappa n_\kappa \left(\mbox{N}_{i \kappa} + \frac{1}{2} \right)~,
\end{equation}
of an equivalent distribution of particles along the different axes
(with ${\rm N}_{i\kappa}$ the partial HO quantum numbers), 
what is called a ``{\it self-consistency condition}$\,$'' in Ref.~\refcite{BM75}.
                                                                    \\[ -2.0ex]

The MI shell correction is thus defined as the difference 
between the quantum GRB MI (\ref{MIGRB}) and its statistical average, 
so that we obtain the total nuclear MI as the sum of a
smooth MI of the
ETF model and this shell correction. 
This way to proceed is perfectly in line with the standard Strutinsky SCM 
where the energy shell correction is defined as a difference between the 
sum of s.p.\ energies and its average, and the nuclear energy is then 
obtained as the sum of the realistic liquid-drop (or ETF) energy and the 
shell correction determined in this way \cite{BD72,BB03}.

\section{Quantum calculations}
\l{sec-qm}

In this section, we will 
describe a system of independent fermions (nucleons)
moving in a deformed mean field of the form of a Woods-Saxon (WS) potential of
spheroidal shape with 0$z$ as symmetry axis.
One then has to solve the Schr\"odinger equation with a potential
\begin{equation}\label{wspot}
 \hspace{-0.2cm}V(r, \theta)=
 \frac{V_0}{1+ \exp\left\{\left[r-R(\theta)\right]/\alpha \right\}}\;,
\end{equation}
where $R(\theta)$ denotes 
in spherical coordinates $\left\{ r,\theta,\varphi \right\}$
the radius of the spheroidal surface and $\alpha$ the 
surface diffuseness.
Introducing semiaxes $a$ and $b$ through the equation
\be\l{spheroid}
  \frac{x^2+y^2}{a^2} + \frac{z^2}{b^2} = 1 \;,
\ee
where, because of volume conservation, one must require that $a^2b=R_0^3$,
with $R_0$ the radius of the corresponding spherical shape, one can define
through 
 $\eta = b/a$ 
a deformation parameter that will be larger one for 
prolate ($b>a$) and smaller one for oblate ($a>b$) shapes.
                                                                    \\[ -2.0ex]

To solve the Schr\"odinger (eigenvalue) equation with the WS 
potential (\ref{wspot}) one can use the expansion of the WS eigenfunctions 
in the basis of a deformed axially-symmetric harmonic-oscillator
(HO)\cite{DP69}, as explained in the appendix. The particle number density 
$\rho(\varrho,z)$ can then be written, in cylindrical coordinates 
$\left\{\varrho,z,\varphi\right\}$ with $\varrho=\sqrt{x^2+y^2}$, in the 
standard form:
\be\l{denho}
 \rho(\varrho,z)=\sum_i n_i \Big|\psi_i(\varrho,z,\varphi)\Big|^2\;,
\ee
where the WS eigenfunctions $\psi_i(\varrho,z,\varphi)$ are given in terms 
of the HO eigenfunctions $\Phi_i$ (see Eq.~(\ref{basfun}) of the appendix). 
For the MI of statistical equilibrium rotations one has 
$\Theta \approx \Theta_{\rm GRB}$, where 
\be\l{GRB}
 \Theta_{\rm GRB}=m\int \d \r\; 
 r^2_\perp \; \rho(\varrho,z) = \sum_i n_i \, \Theta_i
\ee
with
\be\l{Qi}
\Theta_i=m\sum_{j,k}A_{ij} A_{ik} \left(J^{(y)}_{jk}+J^{(z)}_{jk}\right)\;. 
\ee
Here $A_{\mu \nu}$ are the expansion coefficients of the WS eigenfunctions
in the HO basis (see Appendix, Eq.\ (\ref{expHObasis})). 
In Eq.~(\ref{Qi}) we also introduced
\be\l{Jjky}
\!\!J^{(y)}_{jk} \!=\!\! \int\! \d \r\; y^2\; \Phi_{j}^*(\r)\Phi_{k}(\r)
\!=\!\frac{\hbar}{2 m \omega^{}_{\perp}} \delta^{}_{n_z,n_z'}  
\mathcal{Q}^{(y)}_{n_\varrho,n_\varrho'}
\ee
and
\be\l{Jjkz}
\!\!J^{(z)}_{jk}\!=\!\!\int\!\d \r\; z^2\; \Phi_j^*(\r)\;\Phi_k(\r)
\!=\!\frac{\hbar}{m\omega_z}\delta^{}_{n_\varrho,n_\varrho'}\mathcal{Q}^{(z)}_{n_z,n_z'}
\;,
\ee
with respectively
\be\l{calQrnrnrp}
 \mathcal{Q}^{(y)}_{n_\varrho,n_\varrho'} = \int_{0}^{\infty}\xi \d \xi\; \exp(-\xi)\;  
 \mathcal{L}^{(\Lambda)}_{n_\varrho}(\xi)\; \mathcal{L}^{(\Lambda)}_{n_\varrho'}(\xi)
\ee
and 
\be\l{calHnznzp}
\mathcal{Q}^{(z)}_{n_z,n_z'}=\int_{-\infty}^{\infty}\zeta^2\d \zeta\; 
\exp\left(-\zeta^2 \right)\;\mathcal{H}_{n_z}(\zeta)\;\mathcal{H}_{n_z'}(\zeta)\;.
\ee
Finally, these functions are expressed in terms of the
standard Hermite $\mathcal{H}_{n_z}(\zeta)$ and 
associated Laguerre polynomials $\mathcal{L}^{(\Lambda)}_{n_\varrho}(\xi)$ 
(see Eqs.\ (\ref{Lag}) and (\ref{Her}), defined in the dimensionless variables (\ref{arg}) $\xi$ and $\zeta$). 
The calculation of the $\Theta_i$ in Eq.\ (\ref{Qi}) is thus reduced to the 
determination of the transformation matrices $A_{\mu \nu}$ and the calculation
of the simple integrals (\ref{calQrnrnrp}) and (\ref{calHnznzp}),
which can be solved analytically through the recurrence relations of the 
orthogonal polynomials.
To study the correspondence between quantum and classical description 
we will carry out our study with a WS potential (\ref{wspot}) having 
a relatively sharp edge (small diffuseness) and a large depth,
in order to simulate in this way the 
classical motion of particles inside a box of spheroidal shape.

\section{Semiclassical approach}
\l{sec-scl}

Within the POT, both the s.p.\ energy of the system and the MI can be 
subdivided into an average part and a semiclassical shell correction, 
as this has been done for the MI in Eq.\ (\ref{misplit}). It is then 
possible \cite{GM15,MG17} to express these shell components through 
one another
\be\l{dmidE}
  \delta \Theta_{\rm scl} \approx m \, \langle \frac{r_\perp^2}{\vareps} 
                            \rangle^{}_{\rm ETF} \; \delta E_{\rm scl} \; , 
\ee
with a proportionality coefficient given by
\be\l{psav}
 \langle \frac{r_\perp^2}{\vareps} \rangle^{}_{\rm ETF} =
 \frac{\int {\rm d} \vareps\; \vareps\; 
 \int {\rm d} \r {\rm d} \p \; (r_\perp^2/\vareps) \; 
 g^{}_{\rm ETF}(\r,\p;\vareps)}{
 \int \d \vareps\; \vareps 
 \int \d \r \d \p \; g^{}_{\rm ETF}(\r,\p;\vareps)} \; ,
\ee
where $g^{}_{\rm ETF}(\r,\p;\vareps)$ is the ETF approximation to the
semiclassical level-density distribution
$g_{\rm scl}(\r,\p;\vareps) = \partial f^{}_{\rm scl}(\r,\p)/\partial \vareps$
and $f^{}_{\rm scl}(\r,\p)$ 
the Fermi distribution in phase-space \cite{MG17}.
For the simple TF approach, one has
\be\l{distrfuntf}
g_{\rm scl}(\r,\p;\vareps)  \approx
g^{}_{\rm TF}(\r,\p;\vareps) = \delta(\vareps - H(\r,\p)) \;,
\ee
with $H(\r,\p)$ being the classical Hamiltonian.
In the derivation of Eq.~(\ref{dmidE}) for the MI shell 
correction $\delta \Theta_{\rm scl}$ 
the improved stationary phase (periodic orbit) condition
for the evaluation 
of integrals over the phase space variables $\r$ and $\p$ has been
used\cite{MY11,GM15,MG17}.
The main contribution is therefore coming from POs like for the energy shell corrections $\delta E$.
Within the POT,  the PO sum  for the energy shell corrections
$\delta E_{\rm scl}$ writes for an arbitrary potential
well\cite{SM76,SM77,MA02,BB03,MY11,MG17}
\be\l{escscl}
  \delta E_{\rm scl} = d_s\sum_{\rm PO} \frac{\hbar^2}{t_{\rm PO}^2}\,
  \delta g^{}_{\rm PO}(\vareps^{}_F)\;, 
\ee
where $t^{}_{\rm PO} = {\tt M}~t^{{\tt M}=1}_{\rm PO}(\vareps^{}_F)$
is the period of the particle motion along the PO (taking into
account its repetition number ${\tt M}$) and
$t^{{\tt M}=1}_{\rm PO}$ is the period of the particle motion along the 
primitive (${\tt M}=1$) PO,
evaluated at the Fermi energy  $\vareps \!=\! \vareps^{}_F$. 
For the shell correction to the semiclassical level
density (trace formula), one can write
\be\l{dg}
\delta g_{\rm scl}(\vareps)
\cong \sum_{\rm PO} \delta g^{}_{\rm PO}(\vareps)~,
\ee
where
\be\l{densPO}
 \delta g^{}_{\rm PO}(\vareps)= \mathcal{A}_{\rm PO}(\vareps)\;
 \cos\left(\frac{S_{\rm PO}(\vareps)}{\hbar} - 
 \frac{\pi}{2} \mu^{}_{\rm PO}-\phi \right)\;,
\ee
with $\mathcal{A}_{\rm PO}$ being the level-density amplitude. 
In the argument of the cosine function the phase integral $S_{\rm PO}$ 
is the classical action for the PO (or the family of POs),
$\mu^{}_{\rm PO}$ 
the Maslov index (see Refs.~\refcite{BB03,MA17}) and $\phi$ 
an additional phase that depends 
on the dimension of the problem and the degeneracy of the considered orbits \cite{MY11,BB03}.
The Fermi energy $\vareps^{}_F$ has to be
determined by the particle-number conservation condition
(\ref{partconspert}), that can be written in the form
\be\l{partnum}
 N = d_s\int_0^{\vareps^{}_F} \!\! d \vareps\,g(\vareps)\;, 
\ee
where $g(\vareps)$ is the total level density.
                                                                    \\[ -2.0ex]

The rapid convergence of the PO sum in Eq.~(\ref{escscl}) is ensured by 
the factor in front of the density component $\delta g^{}_{\rm PO}$, a 
factor which is inversely proportional to the square of the period of 
motion $t^{}_{\rm PO}$ along the considered PO. 
Only POs with reasonably short periods which occupy a large enough 
phase-space volume will therefore contribute.
Let us mention at this point, that the energy shell correction 
$\delta E_{\rm scl}$ 
that appears 
in Eq.~(\ref{dmidE}) is, of course, through Eq.\ 
(\ref{partnum}), function of the particle number $N$. For the ETF average
$\langle r^2_{\perp} / \vareps \rangle^{}_{\rm ETF}$, Eq.\  (\ref{psav}), 
one can simply use its TF approximation, which 
yields\cite{GM15,MG17}
for the spheroidal cavity an expression through the 
semi-axises $a$ and $b$ of Eq.\ (\ref{spheroid}):
\begin{equation}\label{ell2ms}
  \langle \frac{r_{\perp}^2 }{ \vareps }\rangle^{}_{\rm ETF}
  \approx \langle \frac{r_{\perp}^2 }{ \vareps }\rangle^{}_{\rm TF}
  =\frac{a^2+b^2}{3 \vareps^{}_{\rm F}}\;.
\end{equation}
Expressed in units of the classical (TF), i.e.\ rigid-body MI, 
\begin{equation}\l{RigBody}
 \Theta_{\rm TF}=m \left(a^2+b^2\right) \frac{N}{5} \;,
\end{equation}
one approximately obtains for the MI shell correction (\ref{dmidE})
of the spheroidal cavity
\be\label{dtxtxtf}
 \frac{\delta \Theta_{\rm scl}}{\Theta_{\rm TF}} =
 \frac{5 \, \delta E_{\rm scl}}{3 N \vareps^{}_{\rm F}} \; .
\ee 
In order to verify, within the semiclassical ISPMR  
the adiabaticity condition, $\omega^2 \Theta_{\rm GRB}/2 \ll E(0)$ for the 
calculation of the shell corrections along the yrast line, Eq.~(\ref{Eomeg}),
for the spheroidal cavity, 
one can make use of Eq.\ (\ref{dmidE}) for $\delta \Theta_{\rm scl}$ together 
with the proportionality coefficient of Eq.\ (\ref{ell2ms}) to find
\be\l{adcondspd}
  \omega/\Omega \ll \left[12/\left(\eta^2+1\right)\right]^{1/2},
\ee
where the expression $\hbar\Omega \approx \varepsilon_F/N^{1/3}$
for the distance between major shells has been used. 
This adiabaticity condition is well verified even for large deformations up to 
$\eta \approx 2$ and for very high frequencies 
which, for not too small particle numbers,
correspond to the large angular momentum values of high-spin physics.

Notice that the condition (\ref{adcondspd}) is, in fact, much more
easily obeyed for the same large particle number as compared to the 
classical perturbative condition $\omega/\Omega \ll N^{-1/3}$, because of the small parameter
$N^{2/3}(\omega/\Omega)^2 \ll 1$ introduced
in Ref.\ \refcite{DF04}.

%
\section{Discussion of shell effects} 
\l{sec-comp}

\begin{figure*}[!ht]
\begin{center}
     \includegraphics[width=0.8\textwidth]{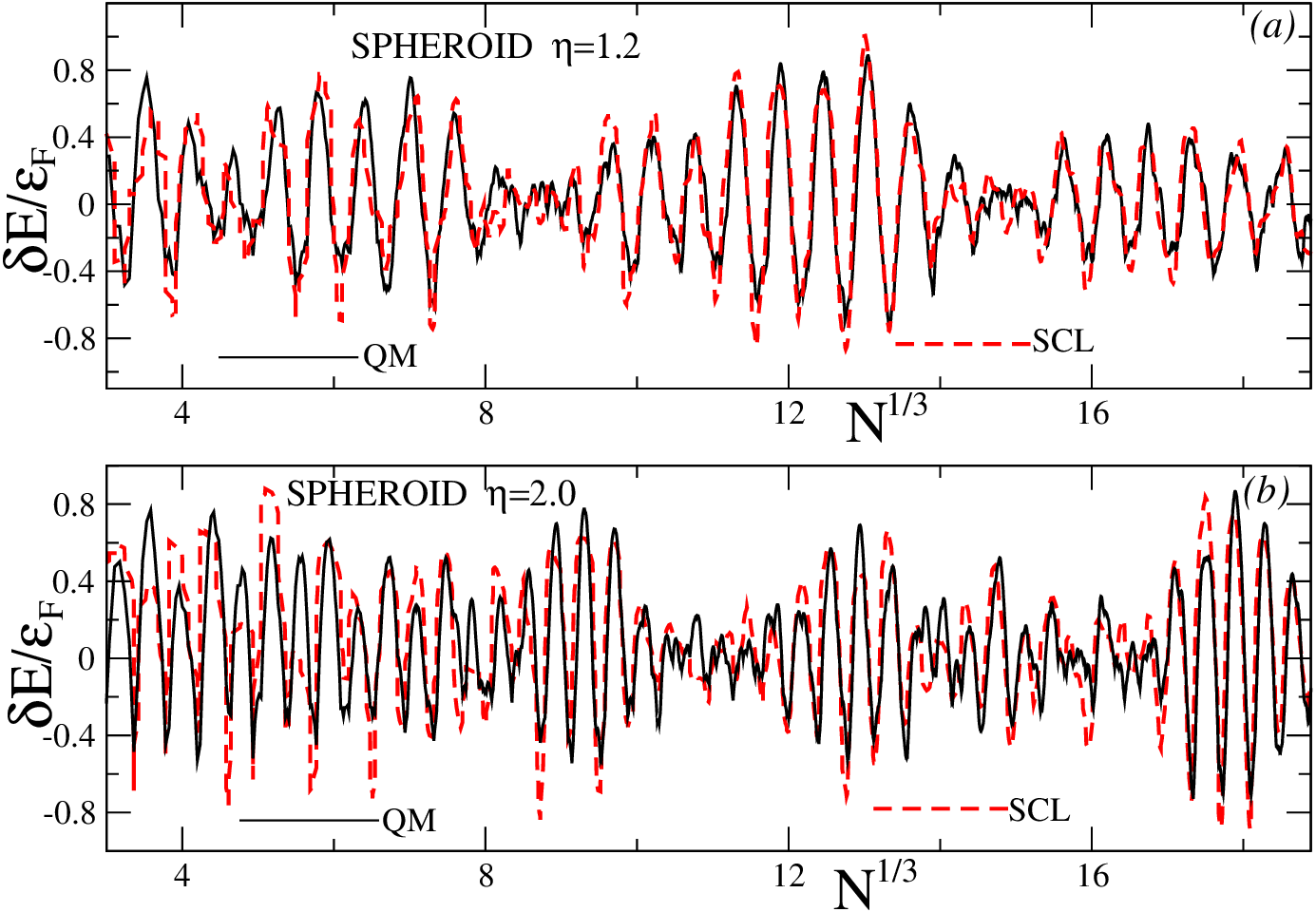} 
\end{center}

\vspace{-0.3cm}
\caption{
{\small 
Quantum-mechanical (solid black) and semiclassical (dashed red line) 
shell-correction 
energies $\delta E$ (in units of the Fermi energy 
$\varepsilon_F$) as function of the cubic root $N^{1/3}$ of the 
particle number, for a spheroidal cavity at 
small $\eta\!=\!1.2$ $(a)$ and large $\eta\!=\!2.0$ $(b)$ 
prolate deformations.
}}
\label{fig1}
\end{figure*}

When calculating the energy shell corrections $\delta E$ for a system of 
$N$ particles in a cavity of spheroidal deformation, one obtains some regular
oscillations as function of $N^{1/3}$ which are presented in Fig.~\ref{fig1} 
for two quite different deformations with axis ratios of respectively
$\eta = 1.2$ and $\eta = 2.0$, i.e.\ for a rather small and for a large 
deformation. This calculation has been carried out, using both the quantum-mechanical (QM) 
and a semiclassical (SCL) resolution of the problem. A solid agreement 
is obtained between both these methods over a very large range of particle 
numbers as shown in the figure, where the energy shell correction is 
displayed in units of the Fermi energy $\varepsilon_F$. In our calculations  
the Fermi energy $\varepsilon_F$ was fixed so as to obtain
\\[4.0ex]
%

\vspace{-0.5cm}
\noindent
 the desired particle number $N$ through Eq.~(\ref{partnum}). It is not astonishing that 
the agreement is less pronounced for small
      particle numbers $N$ where the number of s.p.\ states becomes gradually 
      too small to carry out the Strutinsky smoothing procedure with some 
      reasonable accuracy.
The deep minima (large negative shell corrections) correspond to 
major shell (MS) closures
which are observed in nuclei as well as in metallic clusters.
They are 
for the  here considered potential,
well reproduced in both the quantum and the semiclassical calculations 
and this
as well for small ($\eta = 1.2$) as for large ($\eta = 2$) deformations.
As demonstrated in Refs.\ \citen{MA02} and \citen{MY11},
the factor $1/t_{\rm PO}^2$ in  Eq.\ (\ref{escscl}),
with the time period $t^{}_{\rm PO}$ of the particle motion along the PO,
favours (see also Refs.\ \citen{MS10,MG17,GM15}), 
for the rather small deformation of the 
spheroidal cavity presented in Fig.~\ref{fig1} $(a)$
shorter meridian and equatorial POs 
as already pointed out in the discussion below Eq.\ (\ref{partnum}).
Newborn three-dimensional and hyperbolic orbits, 
on the other hand, are contributing very little at
such a small 
deformation
\cite{MA02,MY11}. 
They, however, become shorter and, hence, 
more important at larger deformations, 
as this is the case in Fig.\ \ref{fig1} $(b)$.
Let us emphasize again that, within 
the ISPM for 
a spheroidal deformation, characterized by $\eta \simg 1.4$, 
the bifurcations of the relatively short equatorial orbits into the
simplest hyperbolic and three-dimensional POs are dominating in the
shell corrections of $\delta E$ and $\delta \Theta$
(see Refs.~\citen{MA02,GM15,MY11}).  
All these properties differ significantly from the results of 
 the classical perturbation theory of
Ref.~\refcite{DF04}, where equatorial orbits do 
not contribute at all and where bifurcating three-dimensional orbits are 
not studied and only suggested among further perspectives.   
                                                    \\[3.0ex]
%
\begin{figure*}[ht!]
\begin{center}   
   \includegraphics[width=0.8\textwidth]{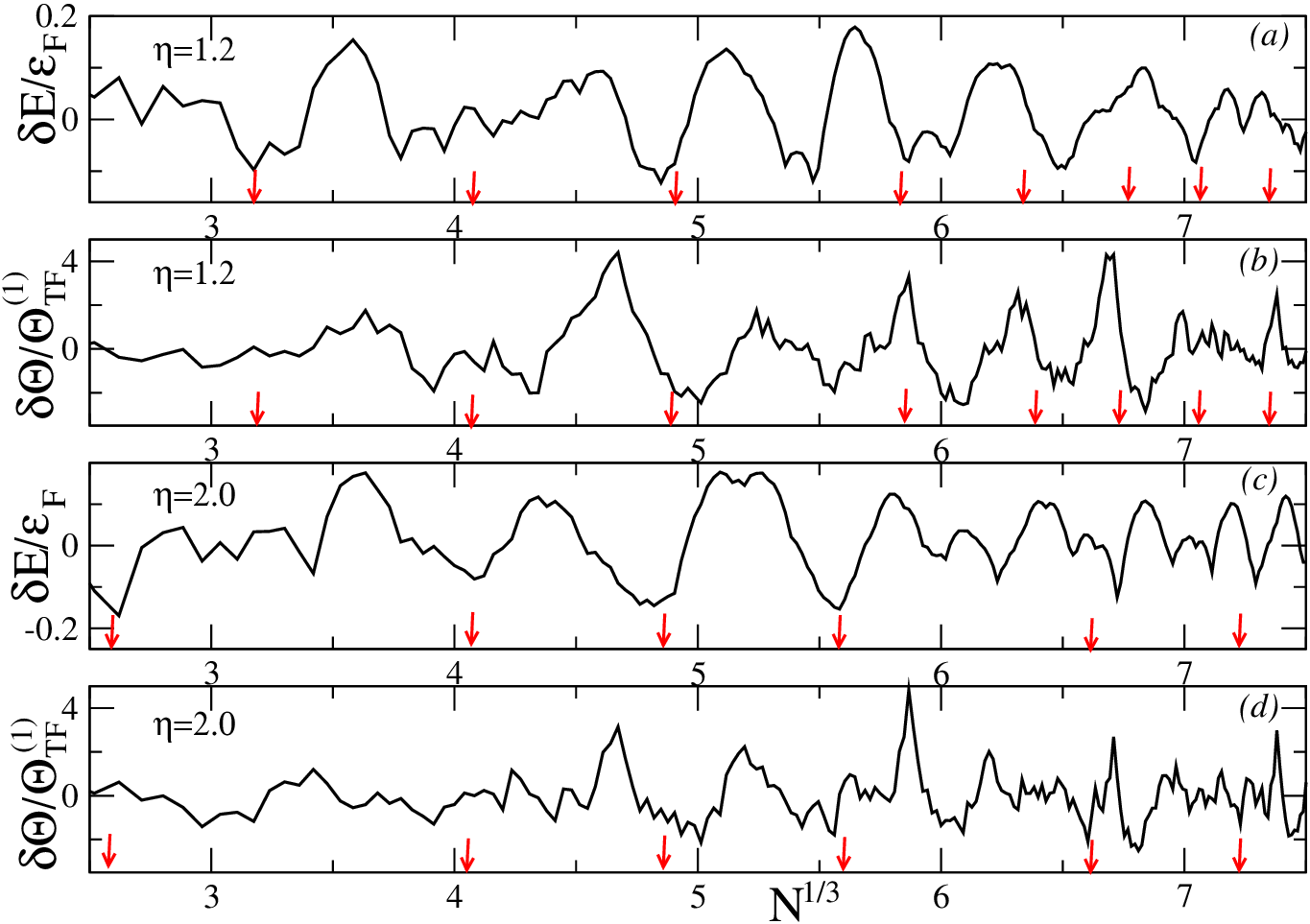}
\end{center}
\caption{
{\small
Quantum-mechanical energy and MI shell corrections,
obtained respectively in units of the
Fermi energy $\varepsilon^{}_F$ and 
of the TF MI per particle $\Theta^{(1)}_{\rm TF}$=$\Theta_{\rm TF}/N$,
with $\Theta_{\rm TF}$ given by Eq.~(\ref{RigBody}), are presented for 
a WS potential (with the depth -300 MeV and diffuseness 0.2 fm, see main text for details) as
functions of the particle number variable $N^{1/3}$, at the same small
($\eta=1.2$ $(a,b)$) and large ($\eta=2.0$ $(c,d)$) deformations 
as for the spheroidal cavity in Fig.~\ref{fig1}. Small red arrows indicate MS
closures, as taken from Fig.\ \ref{fig3} $(c)$ and $(f)$ respectively.  
}
} 
\label{fig2}
\end{figure*}

Fig.\ \ref{fig2} shows a comparison of 
energy and MI shell corrections 
as function of the
particle number
variable $N^{1/3}$, 
for a WS potential of spheroidal shape 
for the deformations $\eta=1.2$ and $\eta=2.0$, respectively. 
This potential is chosen to have a 
constant radius $R_0=r_0\mathcal{N}^{1/3}$ (with $r_0=1.14$ fm), 
for a
fixed
particle number 
$\mathcal{N}=250$ (corresponding 
in the nuclear physics case 
approximately to the center of the Fermium ($Z=100$) 
isotopic chain),
which 
implies
that the radius $R_0$ is fixed
in our calculations to a constant
value $R_0 = 7.18$ fm.
Since in the case of the spheroidal cavity, 
the spectrum is calculated 
in the dimensionless variable $k_i \, R_0$, where 
$k_i=\sqrt{2m \varepsilon_i}/\hbar$ 
with $\vareps_i$ being the energy spectrum
of the cavity,
this dimensionless variable is independent of the specific value 
of the radius $R_0$. One could therefore,
for a comparison of Fig.~\ref{fig2}
with the results of the spheroidal cavity 
(see also Fig.\ \ref{fig1} of Refs.\ \refcite{MA02} and \refcite{GM15}).
formally consider both systems to have the same fixed radius $R_0$.
                                                      \\[ -2.0ex]

The shell components $\delta E$ and $\delta \Theta$ of the energy and the MI, 
shown respectively in panels $(a, c)$ and $(b, d)$ of Fig.~\ref{fig2}, 
are calculated as functions of the nucleon number variable, $N^{1/3}$, 
by the standard SCM, in a WS potential well of depth $V_0=-300$ MeV, 
diffuseness $\alpha=0.2$ fm and fixed radius $R_0=r^{}_0\mathcal{N}^{1/3}=7.18$ fm 
($r^{}_0=1.14$ fm, $\mathcal{N}=250$),
i.e.\ with a fixed s.p.\ spectrum. 
We have chosen a small diffuseness $\alpha$ and a large depth $V_0$ for the
WS-type potential in order to verify
the quantum relationship between $\delta \Theta$ and $\delta E$ for this 
potential, now close to a spheroidal cavity, by comparing it with the semiclassical
relationship, Eq.~(\ref{dmidE}), derived analytically \cite{GM15,MG17}
for any potential well, in particular for the spheroidal cavity.
                                                                    \\[ -2.0ex]

We have found that one is not able to choose
the diffuseness 
smaller than $\alpha \approx 0.2\;$fm 
for that large a WS potential depth 
because then the expansion of the WS eigenfunctions in 
the HO basis states becomes badly convergent, and would require 
a prohibitive number of major shells $n^{}_{\rm HO}$ in the HO spectrum 
to be taken into account. 
The reason why we are choosing such a {\it pathological} form for our 
WS potential is that we want it to resemble as much as possible to the cavity of 
Fig.\ \ref{fig1} to be able to compare the quantum results of the WS potential 
with those obtained analytically in the semiclassical approximation 
for $\delta \Theta $ in the spheroidal cavity. 
      Note also that there is a difference in the plateau
conditions for the calculation of the MI shell corrections $\delta \Theta$
found from the s.p.\ sum in Eq.~(\ref{GRB})  as
compared to the energy shell correction calculation. The reason is that
the quantities $\Theta_i$,
Eq.~(\ref{Qi}), which take the role of a
``s.p. spectrum'' in  Eq.~(\ref{GRB}),  differ
from the 
energy spectrum $\varepsilon_i$ by the coefficients
$A_{\mu \nu}$
from the expansion (\ref{expHObasis}) of the WS 
eigenfunctions
in the 
deformed
HO basis. 
One also finds a relatively significant increase of the number
 $n^{}_{\rm HO}$ of shells in the HO basis that need to be taken into
 account for the calculation of the MI shell component $\delta \Theta$
from 10 to 12 or 14
as compared to the energy shell correction $\delta E$. 
This also leads to somewhat different values of the
Strutinsky smoothing parameters 
for the energy
and the MI shell components. These Strutinsky 
averaging parameters 
for the energy spectrum 
$\varepsilon_i$ are found about the
same for the whole region of particle numbers $N^{1/3}$ shown in
Fig.~\ref{fig2}. 
If one now decreases 
the diffuseness $\alpha$ of the WS potential from a 
value of $\alpha\approx 0.6$ fm, 
realistic for a nuclear
mean field,  
to a value of 
$\alpha\approx 0.2$ fm and simultaneously increases 
the depth of the potential well from about $-50$ MeV to 
$-300$ MeV, one makes that potential resemble a spheroidal cavity. 
Being then able to compare our quantum WS results 
with the result of the semiclassical calculation for that cavity,
one has to find some reasonable choice for $\alpha$ in order to be 
able to still satisfy the Strutinsky plateau condition for the 
average over the energy spectrum. Such a compromise is achieved 
in our calculations presented
in Fig.\ 2, for the parameters of the WS potential well as indicated above, 
at values of the particle number variable $N^{1/3}$ 
between 3 and 7
corresponding to particle numbers in the range
$30 \siml N \siml 340$.
                                                                    \\[ -2.0ex]

The MI shell corrections
in the WS potential, as presented in Fig.\ \ref{fig2} $(b,d)$,
are divided by the TF MI per particle, Eq.~(\ref{RigBody}), that scales with 
the particle number $N$,  $\Theta^{(1)}_{\rm TF}=\Theta_{\rm TF}(N)/N=m(a^2+b^2)/5$.
Here one should take into account that, because of 
using a constant radius $R_0$, this 
TF MI $\Theta_{\rm TF}$ 
is proportional to the particle number $N$.
Therefore, 
when displaying the ratio 
$\delta \Theta / \Theta^{(1)}_{\rm TF}$ as function of $N^{1/3}$,
as we do in Fig.\ 
\ref{fig2}, 
        one obtains 
        a result that has         
essentially
a constant amplitude.
Note that the amplitude of  $\delta E_{\rm scl}/\varepsilon_F$
as function of $N ^{1/3}$ is also 
practically
constant with 
an only slowly decreasing 
spacing between major shells 
(see Fig.~\ref{fig2}$(a,c)$).
One has to note that the family of periodic orbits that gives the main 
contribution  to the semiclassical shell-correction amplitude for the case of 
the  spheroidal cavity is enhanced as compared to the case of the WS potential.
Due to the integrability of the spheroidal cavity,
the symmetry parameter\footnote{The symmetry (or degeneracy)
parameter $\mathcal{K}$ of a family of POs is the number of single-valued 
integrals of a particle motion of fixed energy, which determines in a unique 
way the action integral of the whole family along the PO
\cite{SM76,SM77,MY11,MA02}.}
$\mathcal{K}$
is larger there, for the orbits with highest degeneracy ($\mathcal{K} = 2$), 
as compared to those in the axially-symmetric WS potential, where the orbits 
with highest degeneracy only have $\mathcal{K} = 1$ for the same deformation.
Therefore, the shell-correction amplitudes,
Eq.~(\ref{dmidE}), of both the 
energy and the MI are expected to be enhanced by a
factor N$^{1/6}$ for the 
spheroidal square-well as compared to those for the WS potential
(see also Ref.~\refcite{MY11}).
                                                                    \\[ -2.0ex]

The MI shell correction $\delta \Theta_{\rm scl}$, Eq.\ (\ref{dmidE}), 
turns out to be much smaller than the classical TF (rigid-body) 
component\footnote{Please recall, when comparing the MI shell 
correction $\delta \Theta$ with the TF MI 
$\Theta_{\rm TF}$, Eq.\ (\ref{RigBody}),
that the quantity $\delta \Theta/\Theta^{(1)}_{\rm TF}$ which is 
displayed 
in  Fig.\ \ref{fig2} $(b,d)$ still needs to be divided by
the particle number $N$, in order to  
evaluate 
the relative importance of both these quantities.},
similar to the 
case of the 
energy shell-correction $\delta E$ 
as 
compared to the corresponding TF term.
Taking therefore into account our discussion in relation to 
Eq.\ (\ref{MIGRB}), we should not expect, 
for consistent collective rotations at an equilibrium deformation
(see introduction and subsection \ref{subsec-stateqrot}), 
a large SCM MI correction, $\delta \Theta$,
as compared to the classical rigid-body (TF) value $\Theta_{\rm TF}$.
    However, many important physical phenomena, such as fission isomerism or
    high-spin physics depend 
    dramatically on these shell effects.
Shell effects are also
expected to play a major role for 
the magnetic susceptibility \cite{FK98,RU96}, as a reaction of a system of charged particles to 
a magnetic field, which are expressed by exactly the same type of equations as
we have for the MI.
There, the oscillating (shell) components are going to be relatively enhanced 
 (see e.g.\ Ref.~\refcite{FK98})
as compared to the case of the MI, studied here.
In contrast to the 
classical perturbation POT approach used in Ref.\ \refcite{DF04},
our ISPMR with 
the {\it small parameter} 
$\hbar/S \sim 1/k^{}_FR_0 \sim N^{-1/3}$ of the semiclassical expansion 
results in the adiabatic approximation, 
even at rather large frequencies, Eq.\ (\ref{adcondspd}), 
in a convergence to the exact quantum result which
is the better the larger the particle number $N$. 
 Our ISPMR for 
the MI shell corrections can, of course, 
be applied for larger (non-adiabatic) rotational frequencies and larger deformations, 
as shown in our present study. 
The PO bifurcations 
observed at   
large deformations play here again a dominant role \cite{GM15,MG17}, like 
in the case of the deformed harmonic oscillator\cite{MS10}.
                                                                    \\[ 3.0ex]

\begin{figure*}[!ht]
\begin{center}
\includegraphics[width=0.8\textwidth]{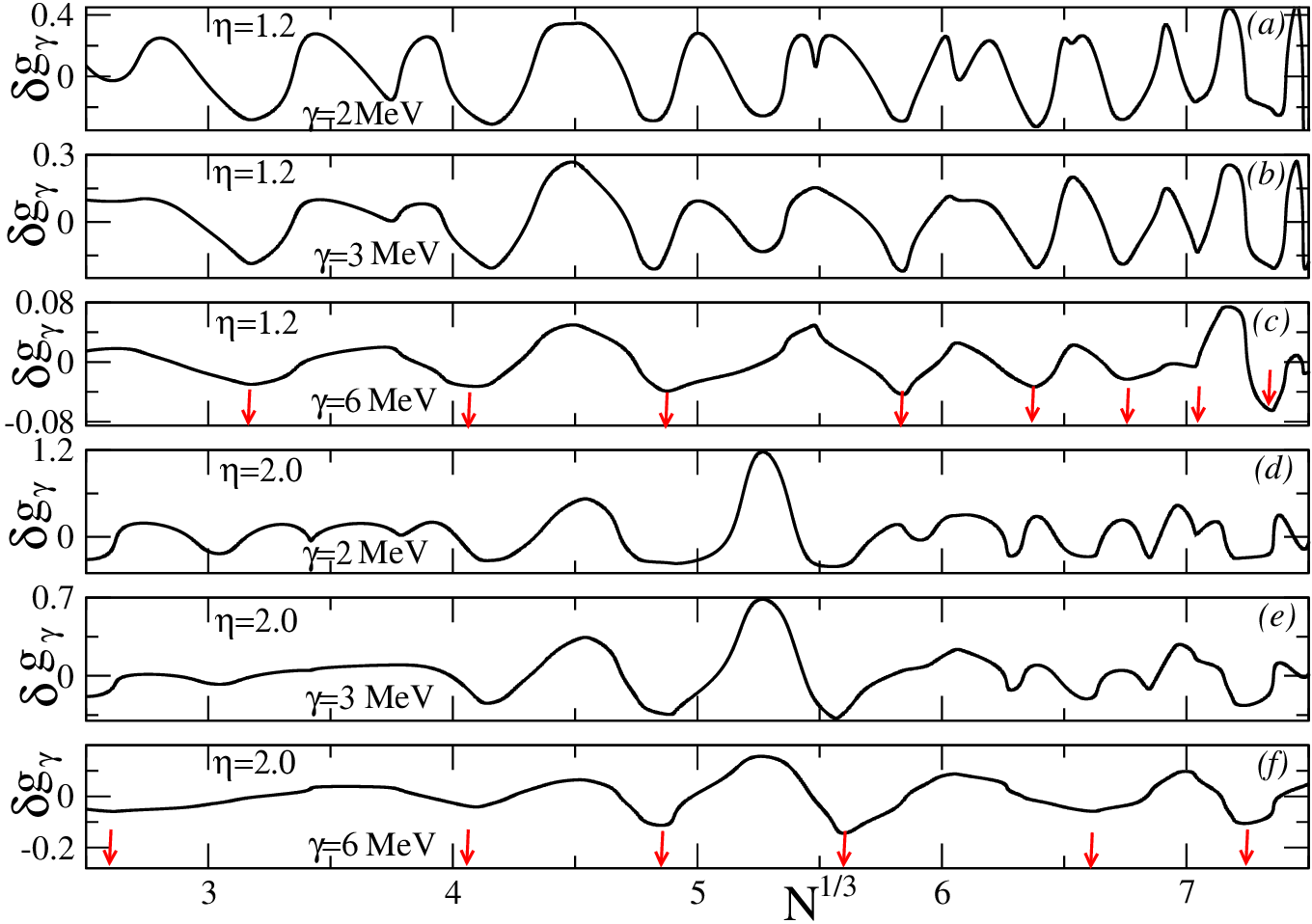}   
\end{center}

\vspace{-0.3cm}
\caption{
  {\small Level-density shell corrections $\delta g_{\gamma}$
(in units of MeV$^{-1}$) as function of the particle number variable $N^{1/3}$
obtained for the WS potential of Fig.\ \ref{fig2} by a Gauss averaging 
with a width parameter $\gamma_{\rm sh}$ as indicated in the
different panels: $(a-c)$ for deformation $\eta\!=\!1.2$,
and $(d-f)$ for $\eta\!=\!2.0$.
Red arrows indicate MS closures. }} 
\vspace{-0.3cm}
\label{fig3}\end{figure*}


In order to be able to test 
over the range of several major shells 
the correspondence $\delta \Theta \propto \delta E~$, Eq.\ (\ref{dmidE}),
in the semiclassical approximation, 
at least what the major shell structure is concerned, 
a large enough interval of particle numbers needs to be considered.
Notice that the semiclassical relationship (\ref{dmidE}) converges 
asymptotically to the quantum results, a convergence which is the
better the larger the particle number $N$, as this can be noticed in Fig.\ 
\ref{fig1}.
On the other hand, when increasing
the number $n^{}_{\rm HO}$ of HO basis shells, one needs  to check the plateau
condition for each value of $n^{}_{\rm HO}$,
and, in fact, for each particle numbers.
This plateau condition for the particle number $N$ is, however, 
much less sensitive for large $N$ values ($N^{1/3} \cong 3-7$).
For the here considered Woods-Saxon potential, two restrictions 
apply to the choice of the particle number N. It cannot be 
chosen too small, otherwise the number of s.p.\ states will not be sufficient
to carry out the Strutinsky smoothing procedure. If, on the other side, the 
particle number becomes too large, s.p.\ states close to the continuum will 
have to be taken into account, which will necessitate a specific treatment
of the continuum,  
which we would like to avoid.
                                                                    \\[ -2.0ex]

To better appreciate the interconnection between the energy 
shell correction $\delta E$ and the corresponding MI shell correction 
$\delta \Theta$, it is important to be able to 
distinguish between major-shell structure (MSS) and corresponding 
sub-shell structure (SSS) 
of the studied $N$ particle system. To this purpose we show 
in Fig.\ \ref{fig3} the level-density shell correction
$\delta g_\gamma(\varepsilon)$ which determines
both of these quantities 
for small ($(a)$ -- $(c)$) and large ($(d)$ -- $(f)$) deformations. 
A Gauss averaging needs to be applied to the level density $g(\varepsilon)$ 
in order to avoid discontinuities 
(of the step-function type) 
in the function $\vareps^{}_F(N)$. 
In order to be able to differentiate between MSS and SSS structures, we need,
however, to use a much larger value of the Gaussian convolution parameter 
$\gamma$ (see Ref.~\refcite{MY11}). 
While a value of $\gamma_0 \approx 0.2$ MeV has to be used in the 
above mentioned 
smoothing of the level density in Eq.\ (\ref{partnum}),
we have used values of $\gamma=\gamma_{\rm sh} \approx 3$ 
for SSS and $\gamma=\gamma_{\rm sh} \approx 6$ for MSS to reveal respectively 
the SS and the MS structures.
This is shown in Fig.\ \ref{fig3}, where MS closures are indicated 
in panels ($c$) and ($f$) through small red arrows. 
It is interesting to 
observe that in the major shell located at $N^{1/3}$ values between 4.9 
and 5.8 (Fig.\ \ref{fig3} $(a)$ and $(c)$) a sub-shell closure can be 
identified for small deformations ($\eta=1.2$) at $N^{1/3}$ around 5.3 
as found in the level-density shell-corrections of Fig.\ \ref{fig3} (a) 
and (b), corresponding to a sub-shell closure for $\delta E$ and $\delta \Theta$
in the vicinity of $N^{1/3} \approx 5.5$ (see Fig.\ \ref{fig2} (a) and (b)).
Similarly, at a large deformations ($\eta=2.0$) such sub-shell structures 
are found between 5.6 and 6.6 in Fig.\ \ref{fig2} $(c)$ and $(d)$. 
For the level-density shell corrections $\delta g^{}_\gamma$, the 
plateau parameters are found to be the 
same as for the energy shell-correction calculations (Fig.~\ref{fig2} ($a,c$)). 
The additional Gaussian width parameter $\gamma_{\rm sh}$ needs 
evidently
to be substantially larger on one hand than the distance between neighboring s.p.\ levels,
but, on the other hand, much smaller than the distance $\hbar \Omega$ between 
major shells, 
so as 
not to wash out completely the main sub-shell or
major shell structures in the here studied range of the particle number variable
$N^{1/3}$ between 3 and 7.
                                                                    \\[ -2.0ex]

Even though the agreement of the variation, as function of the 
particle-number variable $N^{1/3}$, of the shell components 
$\delta E/\varepsilon_F$ and $\delta \Theta/\Theta^{(1)}_{\rm TF}$ 
is not as striking as this was the case for the harmonic oscillator 
potential\cite{MS10}, we still confirm Eq.\ (\ref{dmidE})
in terms of major-shell structures. Indeed,
as this has already been pointed out in the introduction, 
one should not expect to obtain the same quality of agreement,
like a strict proportionality in Eq.~(\ref{dmidE}), between these two 
quantities in a quantum calculation as performed here. 
The simple 
reason is that one is not able then to use the stationary phase approximation 
of the semiclassical approach, valid asymptotically in the limit of large 
particle numbers (see Refs.\ \citen{GM15,MG17}).
Nevertheless, one notices, indeed, the presence of several MS closures 
in the range of $N^{1/3}$ values between 3 and 7 for which a
relatively close correspondence between the energy and the MI shell
correction is observed in terms of major shells.
Note that in the comparison of the quantum shell correction 
$\delta E$ for a smooth-edge WS potential (Fig.~\ref{fig2} ($a$) and ($c$))
with the 
semiclassical result for an infinitely deep spheroidal square-well potential
(see Fig.\ \ref{fig1} and Ref.~\refcite{GM15}),
one needs to take into account different boundary conditions. Indeed, these 
lead to an additional shift of the Maslov 
    phase \cite{BB03,MY11,MA17} in $\delta E$,
    Eqs.~(\ref{escscl}--\ref{densPO}), for the spheroidal cavity
with respect to the 
WS potential, as suggested by the small shift
discrepancy of the last two major shells in 
Fig.~\ref{fig2} 
($a-d$).
Another interesting and nontrivial example is the MS structure 
located, in the $N^{1/3}$ variable, between 4 and 5 for $\eta=1.2$,
and between 2.6 and 4.1 for  $\eta=2.0$, which has respectively a
different SSS in $\delta E$ (see Fig.\ \ref{fig2} $(a)$ or $(c)$) and 
in $\delta \Theta $ (see Fig.\ \ref{fig2} $(b)$ or $(d)$).
                                                                    \\[ -2.0ex]

As becomes evident from Sec.\ 4
a qualitative agreement can be observed between the semiclassical POT 
and the quantum results for both the deformed spheroidal cavity and a deep
and almost sharp-edged WS potential.
We are thus able to establish a {\it statistical correspondence}
of our relation (\ref{dmidE}) between the shell contributions $\delta \Theta$ 
and $\delta E$ on the level of the resolution of major-shell structures,
as this was already observed more strictly for the harmonic oscillator
potential 
\cite{MS10}.

%
\section{Conclusions}
\label{sec-concl}

The shell corrections to the moment of inertia are determined through  
the generalized rigid-body MI for equilibrium rotations beyond the 
quantum perturbation expansion. 
We have shown that, for a WS potential of both small and large spheroidal 
deformation, the semiclassical relation (\ref{dmidE}) between energy and 
moment-of-inertia shell corrections approximately  holds for the major 
shell structure and is in qualitative agreement with the quantum result.
The WS mean-field potential that we have used, has been chosen to have a
large depth and a small surface diffuseness to make it resemble as much 
as possible a square-well mean field. 
In contrast to the cavity, we are thus defining a non fully 
integrable Hamiltonian, in order to test whether 
  our ISPMR  
works also for such a system. 
It is of particular interest to carry out this kind of investigation in a
large range of deformations, since at larger deformations
the bifurcation phenomenon plays an important
role\cite{MA02,MY11,MG17}.
Our study confirms the fundamental property, common in general to all 
finite Fermi systems, and at any deformation, that the ``major-shell 
structure'' of $\delta \Theta$ is related to the one of the energy 
shell corrections $\delta E$ through the inhomogenuity of the 
distribution of single-particle states near the Fermi surface.
A more systematic investigation of the relationship between these two 
shell corrections and a comparison between semiclassical and quantum results
is on our agenda.
                                                                    \\[ -2.0ex]

One of the most important generalizations and
most attractive applications
of the semiclassical periodic-orbit theory seems, however, the inclusion 
into our description of the spin-orbit and the pairing interactions 
\cite{BQ81,SG89,TH95,AB02,BA04,BR10},
and the study of their influence on the 
collective vibrational and rotational excitations in heavy deformed 
nuclei. 
It is, indeed, well known that, in particular, the inclusion of the 
pairing correlations often reduces the MI considerably. Any realistic
description of rotation in nuclei therefore absolutely  
requires to take the pairing degree of freedom into account.
Even if the MI shell component can be small as compared to its
semiclassical counterpart, it is well-known (see e.g.\ Ref.\ \refcite{Fr12})
that e.g.\ for high-spin physics both the average and the 
shell-correction components are important in the non adiabatic 
approximation.
To carry out a full-fledged realistic calculation including a possible 
comparison with experimental data is of course beyond the scope of our
present study, which should be rather understood as a first step into 
this direction. 
It is, in particular far away from the magic numbers, where shell 
corrections are small and pairing correlations are important that such 
a comparison of our theoretical predictions with the experimental data 
on rotational bands in well deformed nuclei would be highly desirable.
                                                                    \\[ -2.0ex]

Now that the applicability of 
  our ISPMR to 
non integrable systems
such as the Woods-Saxon mean field has been demonstrated, we plan of 
course to apply our method to nuclear systems with a more realistic 
surface diffuseness 
within the nuclear collective dynamics, 
in particular involving magic nuclei, where the above discussed effects 
should be strongest.
In this connection it is obvious that 
  our ISPMR results 
could also 
be extremely interesting for the calculation of shell effects 
of the magnetic susceptibility in quantum dots \cite{FK98,RU96}.

\section*{Acknowledgements}

The authors gratefully acknowledge J.P.\ Blocki and A.I.\ Sanzhur for many fruitful
discussions. We are also very grateful for many creative discussions
with K.\ Arita, R.K.\ Bhaduri, M.\ Brack, S.N.\ Fedotkin,
S.\ Frauendorf, A.N.\ Gorbachenko, F.A.\ Ivanyuk, V.M.\ Kolomietz,
M.\ Matsuo, K.\ Matsuyanagi, and V.A.\ Plujko.
One of us (A.G.M.) is very grateful for the nice hospitality extended
to him during his working visits of the National Centre for Nuclear
Research in Otwock-Swierk/Warsaw, the Hubert Curien Institute of the
Strasbourg University, the University of Regensburg/Germany, and the
Physics Department of the Nagoya Institute of Technology/Japan.
Many thanks also go to the Japanese Society for the Promotion of
Sciences for their financial support, Grant No. S-14130.
This work was also supported by the budget program ``Support for the
development of priority areas of scientific research'' of the National
Academy of Sciences of Ukraine (Code 6541230, No. 0120U100434).

\appendix

\section{Harmonic oscillator basis}
\l{appA}

For the quantum calculations of the MI and the energy shell corrections in the
spheroidal WS potential (\ref{wspot}), one can perform the diagonalization
procedure through an expansion of the WS eigenfunctions $\psi_i(\r)$ in the 
basis of a deformed harmonic oscillator \cite{DP69}:
\be\l{expHObasis}
 \psi_i(\r) = \sum_{j} A_{ij} \; \Phi_{j}(\r) \;.
\ee
The HO basis states $\Phi_{j}$ 
are defined in cylindrical coordinates 
$\left\{\varrho,z,\varphi\right\}$ 
as
\be\l{basfun}
 \Phi^{}_j(\r) = \vert j \rangle = \vert n_z n_\varrho \Lambda \rangle 
 = \mathcal{R}_{n_\varrho}^{(\Lambda)}(\varrho)\;
\mathcal{Z}^{}_{n_z}\!(z) \;  \phi^{}_{\Lambda}(\varphi) \;, 
\ee
where $n_z, n_\varrho,$ and $ \Lambda$ are the quantum numbers of the state
and
\be\l{Rfun}
\mathcal{R}_{n_\varrho}^{(\Lambda)}(\varrho)=
\left(\frac{2 m \omega^{}_{\perp}}{\hbar}\right)^{1/2}\; \exp(-\xi/2)\;
\mathcal{L}^{(\Lambda)}_{n_\varrho}(\xi)\;,
\ee
\be\l{Zfun}
 \mathcal{Z}_{n_z}(z)= \left(\frac{m \omega_z}{\hbar}\right)^{1/4} \;
                         \exp(-\zeta^2/2)\; \mathcal{H}^{}_{n_z}(\zeta)\;,
\ee
\be\l{Ffun}
\phi^{}_{\Lambda}(\varphi)=(2 \pi)^{-1/2}\;\exp(i \Lambda \varphi)\;,
\ee
with
\be\l{arg}
  \xi^{1/2} = \sqrt{\frac{m \omega^{}_\perp}{\hbar}} \varrho = b^{}_\perp \, \varrho \;, 
  \quad
  \zeta = \sqrt{\frac{m \omega_z}{\hbar}} z = b^{}_z \, z \;. 
\ee
The frequencies $\omega_\perp$ and $\omega_z$ of the 
axially-symmetric HO basis
are connected, as usual, 
by the volume conservation
condition $\omega_\perp^2\omega_z=\omega^3_0$, with 
$\omega_\perp/\omega_z=q$ 
being the deformation parameter of the basis, i.e.\ 
$\omega_\perp=\omega^{}_0q^{1/3}$ and $\omega_z=\omega^{}_0q^{-2/3}$. 
It is convenient to use dimensionless  
      variables 
as we have done through Eqs.\ (\ref{arg}) by introducing an 
inverse length $b^{}_0=\sqrt{m\omega^{}_0/\hbar}$
as a parameter of the HO basis (for  
nuclear systems with $\mathcal{N} \sim 200-300$ considered in our study,
one obtains together with 
$\hbar \omega^{}_0 \approx 50\,$MeV$/\mathcal{N}^{1/3}$ 
a value 
of $b^{}_0 \approx 0.45\,$fm$^{-1}$)
and consequently corresponding inverse lengths 
$b^{}_\perp = b^{}_0 \, q^{1/6}$ and $b^{}_z = b^{}_0 \, q^{-1/3}$. 
The functions $\mathcal{L}^{(\Lambda)}_{n_\varrho}(x)$ and
$\mathcal{H}^{}_{n_z}(x)$ 
in Eqs.\ (\ref{Rfun}) and (\ref{Zfun})
are related 
to the standard generalized Laguerre $L_n^{(\Lambda)}(x)$, 
and Hermite $H_n(x)$ polynomials by 
\be\l{Lag}
\mathcal{L}^{(\Lambda)}_n(x) = \left(\frac{n!}{(n+\Lambda)!}\right)^{\!\!1/2}
       \!\! x_{}^{\Lambda/2} \; L_n^{(\Lambda)}(x)\;,
       \ee
and
\be\l{Her}
\mathcal{H}_n(x)=(2^n n!\pi^{1/2})^{-1/2}\;H_n(x) \;.
\ee
The functions $\mathcal{L}_{n}^{\Lambda}(x)$ and $\mathcal{H}_{n}(x)$ 
obey orthogonality relations similar, up to constants, to those of
the Laguerre, $ L_n^{(\Lambda)}(x)$, and Hermite, $H_n(x)$, polynomials 
themselves \cite{DP69}.

%
%


\end{document}